
\input amstex
\documentstyle{amsppt}
\NoRunningHeads
\NoBlackBoxes
\hyphenation{Ko-lo-kol-tsov}
\magnification=1200
\document
\bf\ \newline
BELAVKIN-KOLOKOLTSOV WATCH--DOG EFFECTS IN INTE\-RAC\-TIVELY CONTROLLED
STOCHASTIC COMPUTER-GRAPHIC DYNAMIC SYSTEMS.
A MATHEMATICAL STUDY

\

\

\rm Denis V. Juriev

\

\ \newline
Mathematical Division, Research Institute for System Studies
[Information Technologies], Russian Academy of Sciences, Moscow, Russia
(e mail: juriev\@systud.msk.su)

\

\ \newline
Laboratoire de Physique Th\'eorique de l'\'Ecole Normale Sup\'erieure,
24 rue Lhomond, 75231 Paris Cedex 05, France
(e-mail: juriev\@physique.ens.fr)

\

\

\

\

{\bf Abstract.} Stochastic properties of the long time behavior of
a continuously observed (and interactively controlled) quantum--field top
are investigated mathematically. Applications to interactively controlled
stochastic computer-graphic dynamic systems are discussed.
\newpage

\head I. INTRODUCTION\linebreak (DESCRIPTION OF PROBLEMS, MOTIVATIONS AND
GENERAL DISCUSSIONS)\endhead

The main difficulty to account the high--frequency eye tremor in {\it
mobilevision\/} ({\it MV\/}) (Juriev 1992, 1994a, 1994b) is that in this case
a solution of the complete MV evolution equations in real time requests about
10$^8$--10$^9$ arithmetical operations per second (moreover, it needs special
displays of a high refreshing rate ($\sim$ 300--500 frames per second) and a
small image inertia). Such account may be performed only on a narrow class of
computers for the purposes of scientific experiments on peculiarities of human
vision in interactive computer-graphic systems (Juriev 1992, 1994a), but it is
very inconvenient for an assimilation of MV as a computer-graphic tool f.e.
for an interactive visualization of 2D quantum field theory (Juriev 1994c). So
one should to use some stochastic simulation of the interactive processes,
i.e. to consider an imitated stochastic process instead of the tremor. Such
approach leads to {\it stochastic mobilevision\/} ({\it SMV\/}) (Juriev 1992),
which evolution equations have a stochastic Belavkin--type form (Belavkin
1988, Belavkin \&\ Kolokoltsov 1991, Kolokoltsov 1991). It seems that the
interactive effects for ordinary MV and SMV are similar in general, because
the interactive processes accounting saccads are not stochastized; though it
is not an undisputable fact that they are always identical (f.e. in a
situation of the so--called "lateral vision"). The combination of MV with
cluster and spline techniques allows to work on computers with 10$^5$--10$^6$
arithmetical operations per second (as well as to use simpler devices for eye
motion detection and a wider class of displays), whereas all enumerated above
circumstances make the tremor accounting in terms of ordinary MV almost
unreasonable nowadays.

Nevertheless, all these advantages of SMV are not crucial in view of the
permanent progress in the computer hightech (for example, the using of a
distributed parallel processing allows to diminish the request for the tremor
accounting ordinary MV to $\sim$ 10$^6$ arithmetical operations per second,
etc.). A deeper advantage of SMV is more theoretical --- it is a presence of
the Belavkin--Kolokoltsov watch--dog effects (Kolokoltsov 1993, see also the
original papers (Chiu {\it et al\/} 1977, Misra \&\ Sudarshan 1977), where
"watch--dog effects" or a "quantum Zeno paradox" were put under a
consideration, and a recent note (Home \&\ Whitaker 1992) for references and a
description of a current state of the problem, or the book (Peres 1993) for a
general point of view) in SMV in certain rather natural and general cases (i.e.
for certain values of internal parameters measuring the degree of localization
of interaction) that means an {\it a priori\/} finiteness of
sizes of stochastic "cores" of an image during observation, moreover they may
be diminished to several pixels by a suitable choice of a {\it free\/}
controlling parameter (the so--called {\it "accuracy of measurement"\/}
(Belavkin 1988, Belavkin \&\ Kolokoltsov 1991, Kolokoltsov 1991, 1993)). The
watch--dog effect may be considered as a weaker but also tamer form of
nondemolition than the quasistationarity (Juriev 1992, 1994a): there exists a
wide class of models, in which the first is observed whereas the least is
broken, one may consider canonical projective $G$--hypermultiplets (Juriev
1994a) (see also Juriev 1994b) as a simple example.

Thus, a transition from MV to SMV partially solves {\it a priori\/} the main
problem of dynamics in interactive psychoinformation computer-graphic systems
(Juriev 1992, 1994a) --- a problem of the nondemolition of images by the
interactive processes (i.e. their stability under observation). Certainly, SMV
does not solve the nondemolition problem completely {\it a priori}. It only
garantees that the stochastic cores of image have finite sizes during
observation, it means that details of image do not diffuse. Nevertheless,
they may move, being ruled by the slow eye movements. So though details of
image are perserved, the image may be destructed as a whole. It seems that the
quasistationarity conditions (Juriev 1992, 1994a) are realistic complements to
watch--dog effects and together they provide a complete long--time
nondemolition of images.

Also it should be marked that such {\it a priori\/} nondemolition in SMV
confirms a presence of {\it a posteriori\/} one in the tremor accounting
ordinary MV.

The purpose of this note is to investigate the Belavkin--Kolokoltsov
watch--dog effects in SMV mathematically.

Summarizing arguments above one may conclude that such investigations are
motivated by the overlapping of two problems:
\roster
\item"1)" the difficulty to account the high--frequency eye tremor in ordinary
mobilevision, which leads to the necessity to consider tremor's stochastic
simulations;
\item"2)" the main problem of dynamics in interactive psychoinformation
computer-graphic systems, i.e. a problem of the nondemolition of images by the
interactive processes; it motivates investigations of long--time properties of
(nonlinear) stochastic dynamics in SMV.
\endroster
So the first problem explains, why stochastic mobilevision is put under a
consideration, the second one explains a choice of questions, which are
tried to be solved in the paper.

\head II. MATHEMATICAL SET UP\linebreak (DEFINITION AND COMMENTS) \endhead

First of all, stochastic mobilevison as well as ordinary mobilevision are
interactive computer-graphic systems, the evolution of images in which is
governed by the eye movements in accordance to the certain {\it dynamical
perspective laws}, i.e. dynamical equations, which govern an evolution of
image during observation (see Juriev 1994b,c). So their definitions are just
the specifications of such laws (it should be specially stressed that we
restrict now our interest in interactive computer-graphic systems by an
intrinsic {\it constructive\/} point of view (cf. Kaneko \&\ Tsuda 1994),
considering them {\it as such\/} but not as {\it descriptive\/} tools of any
use for modelling or visualizing of various physical processes (as in Juriev
1994c), such approach may be rather narrow but effective and it is reasonable
to adopt it for the further discussion). The laws for MV were written in
(Juriev 1992, 1994a,b,c). Stochastic mobilevision have the slightly different
laws. A difference may be briefly summarized in the following terms: (1) the
high--frequency eye tremor is decoupled from the slow eye motions (including
saccads), (2) it is stochastized in such a way that it may be considered as
{\it purely internal process\/} in the system so that (3) its characteristics
are not completely determined by the real eye motion and may be reinforced.

This qualitative description of stochastic mobilevision is sufficient for the
understanding of results as well as their significance for applications but we
need in a more formal definition for their deduction. However, a reader, which
is not interested in formal expositions may omit all mathematical constuctions
below and restrict himself to some comments.

Note once more that to define stochastic mobilevision means to specify its
dynamical perspective laws (dynamical equations, which govern an evolution of
image during observation) and we prefer to do it rather formally in purely
mathematical terms. Such specification is rather analogous to one for the
ordinary MV and is based on concepts of 2D quantum field theory. However, it
should be noted that this paper is not a suitable place for a new review of
mathematical foundations of MV (and, certainly, of 2D QFT), the detailed
exposition of which may be found in the papers (Juriev 1994b,c) (see also
original papers (Juriev 1992, 1994a)). So a knowledge of mathematical
formalism for ordinary mobilevision is presupposed. Certainly, all necessary
objects will be formally introduced below, but many motivations for
definitions, constructions and notations as well as unavoidable remarks on
their interrelations are omitted if they are just the same as for ordinary
MV. In particular, all objects of 2D QFT are used without any comments. It is
explained by the fact that a detailed presentation of all related (sometimes,
rather mathematically technical) material would rather overload an exposition
and would not help to the clarification of general ideas and the understanding
of results. So it should be emphasized once more that the description of
stochastic mobilevision will be rather formal, whereas the interpretations of
mathematical results and their significance for applications will be commented
in detail throughout the text, in the conclusion and in remarks on
applications after it.

\definition{Definition} Let $H$ be a canonical projective
$G$--hypermultiplet (Juriev 1994a,b), $A_t(u,\dot u)$ -- an angular field
(obeying the Euler--Arnold equations $\dot A_t=\{\Cal H,A_t\}$, where the
hamiltonian $\Cal H\in S^{\cdot}(\frak g)$ ($\frak g$ is the Lie algebra of a
Lie group $G$) is a solution of the Virasoro master
equation) (or its finite--dimensional lattice approximations (Juriev 1994c)).
Let $J(u)$ --- an additional $q_R$--affine current (Juriev 1994b,c)(or its
finite--dimensional lattice approximation (Juriev 1994c)) commuting with $G$.
A stochastic evolution equation $$d\Phi(t,[\omega])=A_t(u,\dot
u)\Phi(t,[\omega])\,dt+\lambda J(u)\Phi(t,[\omega])\,d\omega,$$ where
$d\omega$ is the stochastic differential of a Brownian motion (i.e.
$\frac{d\omega}{dt}$ is a white noise), will be called {\it
the\/} ({\it quantum--field\/}) {\it Euler--Belavkin--Kolokoltsov formulas},
the parameter $\lambda$ will be called {\it the accuracy of measurement\/}
(cf. Belavkin 1988, Belavkin \&\ Kolokoltsov 1991, Kolokoltsov 1991).
\enddefinition

\remark{Remark 1} These formulas are a reduced version of more general ones
$$d\Phi(t,[\omega])=\{A_t(u,\dot u)+\alpha\lambda^2\,:\!J^2(u)\!:\,\}
\Phi(t,[\omega])\,dt+\lambda J(u)\Phi(t,[\omega])\,d\omega,$$ which will be
also called {\it the\/} ({\it quantum--field\/}) {\it
Euler--Belavkin--Kolokoltsov formulas}; $\lambda^2:\!J^2(u)\!:$ is a
Belavkin--type quantum--field counterterm (cf. Belavkin 1988, Belavkin \&\
Kolokoltsov 1991, Kolokoltsov 1991), where $:\!J^2(u)\!:$ is a
spin--2 primary field received from the current $J(u)$ by the truncated
Sugawara construction (Juriev 1994a).
\endremark

Here $u=u(t)$ and $\dot u=\dot u(t)$ are the slow variables (Juriev 1992) of
observation (sight fixing point and its relative velocity), the tremor is
simulated by a stochastic differential $d\omega$, $\lambda$ is a {\it free\/}
parameter, $\Phi=\Phi(t,[\omega])\in H$ is a collective notation for a set of
all continuously distributed characteristics of image (Juriev 1992, 1994b,c),
$q_R$ is a free internal parameter of a model, which measures the degree of
localization of interaction (the local case corresponds to $q_R=0$). The most
important case is one of $q_R\ll 1$ and all our results will hold for this
region of values of $q_R$.
The stochastic Euler--Belavkin--Kolokoltsov formulas coupled with the
deterministic Euler--Arnold equations define a dynamics, which may be
considered as {\it a candidate\/} for one of {\it a continuously observed\/}
({\it and interactively controlled\/}) {\it quantum--field top} (Juriev
1994a).

\remark{Remark 2} It should be specially emphasized that in stochastic
mobilevision $\lambda$ is a {\it free\/} parameter, which may be chosen
arbitrary by hands (f.e. as great as it is necessary). It means that slow
movements (including saccads) and tremor are decoupled, the firsts are
considered such as in an ordinary MV, whereas the least is stochastized in a
way that {\it its amplitude may be reinforced}.
\endremark

\remark{Remark 3} As it was mentioned above the internal parameter $q_R$
measures a degree of localization of a man--machine interaction in MV and SMV.
It is natural to suppose that the Belavkin--Kolokoltsov watch--dog effects
will appear for sufficiently small values of $q_R$ and the condition $q_R\to 0$
will produce the diminishing of stochastic cores of image. Indeed, we shall see
that sizes of stochastic cores diminish if $q_R$ tends to $0$ and $\lambda$
increases.
\endremark

Below we shall work presumably with finite--dimensional lattice approximations
(cf. Kolokoltsov 1993) and the associate evolution equation in $H^*$ (Juriev
1994c), keeping all notations. Also $\Phi$ will be considered as defined on a
compact (the screen of a display or a cluster). It should be marked that in
this case the Euler--Belavkin--Kolokoltsov formulas are transformed into the
ordinary (matrix) stochastic differential equations of diffusion type (Gihman
\&\ Skorohod 1979, Skorohod 1982), and hence, $\Phi=\Phi_t=\Phi(t,[\omega])$
is a diffusion Markov process (Dynkin 1965).

\remark{Remark 4} Lattice approximations of the ordinary (unobserved and
non--controlled) quantum--field top (in this case angular fields are reduced
to single currents) were actively investigated by St.Petersburg Group directed
by Acad.L.D.Faddeev (Alekseev {\it et al\/} 1991, 1992). The main difficulties
(technical as well as principal) in their treatments were caused by a locality
of ordinary ($q_R=0$) affine currents. However, $q_R$--affine currents are not
local so their discretizing is easily performed (Juriev 1994c). It is very
interesting to receive lattice current algebras of (Alekseev {\it et al\/}
1991, 1992) from naturally discretized $q_R$--affine currents by a limit
transition $q_R\to 0$, but this problem is a bit out of the line here.
\endremark

The fact that the ordinary quantum--filed top may be received as a particular
case of our construction ($\lambda=0$, $q_R=0$, $A(u,\dot u)=J(u)\dot u$,
where $J(u)$ is a current) motivates to consider our object as a continuously
observed (and interactively controlled) quantum--field top. Continuous
observation means the inclusion of a stochastic term ($\lambda\ne 0$), whereas
the interactive controlling means the presence of complete algular fields
$A(u,\dot u)=\sum_k B_k(u)\dot u^k$, where $B_k(u)$ are primary fields of spin
$k$, instead of single currents. It seems that these arguments are sufficient
for our terminological innovation.

\remark{Remark 5} The Euler--Belavkin--Kolokoltsov formulas are {\it
postulated\/} to be the dynamical perspective laws for stochastic mobilevision.
So they are regarded as {\it a mathematical definition of SMV\/} (cf. Juriev
1994b,c). From such point of view a transition from MV to SMV consists in:
\roster
\item"1)" the decoupling of slow movements (including saccads) and tremor;
\item"2)" a stochastization of tremor;
\item"3)" the setting the controlling parameter $\lambda$ free, so that its
value may be chosen by hands and it is not completely determined by real
parameters of the eye motions.
\endroster
Thus, the main difference between MV and SMV is that tremor in MV is {\it an
external process\/} governing an evolution of a computer graphic picture,
whereas
its stochastization is {\it an internal process\/} (in spirit of {\it
endophysics\/} of O.E.R\"ossler (R\"ossler 1987)) and its characteristics may
be specified by hands.
\endremark

Let's summarize the material of this paragraph. Note once more that the
ordinary mobilevision is an interactive computer-graphic system, the evolution
of images in which is governed by the eye movements in accordance to the
certain dynamical perspective laws, which were written in (Juriev 1992,
1994a,b,c). Stochastic mobilevision is an analogous interactive
computer-graphic system, but with slightly different dynamical perspective
laws. Namely, in the dynamical perspective laws of MV the high--frequency eye
tremor is decoupled from the slow eye motions (including saccads), is
stochastized in such a way that it may be considered as {\it purely internal
process\/} in the system so that its characteristics are not completely
determined by eye motions and may be reinforced. So the parameters of an
external real process (eye tremor) may be transformed and scaled up to receive
ones an internal virtual process (stochastization of tremor). For the
understanding of results the explicit form of dynamical perspective laws is
not necessary though it is, of course, unavoidable for their deduction, which
is presented in the following paragraph, which may be omitted by a reader
interested only in applications, who may restrict himself by the comment and
remark at its end.

\head III. MATHEMATICAL ANALYSIS\linebreak (THE MAIN STATEMENTS AND
DISCUSSIONS)
\endhead

Let $D_A(\Phi)=\left<A^2-\left<A\right>^2_\Phi\right>_\Phi$,
$\left<A\right>_\Phi=\frac{(A\Phi,\Phi)}{(\Phi,\Phi)}$ (Kolokoltsov 1993). It
should be mentioned that one may consider the Euler--Belavkin--Kolokoltsov
formulas with a redefined quantum field $\tilde J(u)=J(u)-\left<J(u)\right>$
instead of the $q_R$--affine current $J(u)$ to receive a full likeness to the
original Belavkin quantum filtering equation (Belavkin 1988, Belavkin \&\
Kolokoltsov 1991, Kolokoltsov 1991, 1993) if the inner (scalar) product
$(\cdot,\cdot)$ is claimed to be translation invariant and scaling
homogeneous. $E_\Phi$ is the mathematical mean with respect to the standard
Wiener measure for observation process with initial point $\Phi$ (Kolokoltsov
1993).

\proclaim{Lemma 1}
$$(\forall\Phi_0)\quad\limsup_{t\to\infty}
E_{\Phi_0} D_J(\Phi(t,[\omega]))=
K\lambda^{-2}\longrightarrow_{\lambda\to\infty}0.$$
\endproclaim

The l.h.s. expression (multipled by $\lambda^2$, i.e. just the constant $K$)
is called {\it the Kolokoltsov coefficient\/} of quality of measurement
(Kolokoltsov 1993).

\demo{Sketch of the proof} Indeed
$$\aligned\lambda^2\limsup_{t\to\infty}E_{\phi_0}D_J(\Phi(t,[\omega]))=&
\limsup_{t\to\infty}E_{\Phi_0}D_{\lambda J}(\Phi(t,[\omega]))=\\
&\limsup_{t\to\infty}E_{\widetilde\Phi_0}D_J(\widetilde\Phi(t,[\omega])),
\endaligned$$ where $\widetilde\Phi$ is a solution of the
Euler--Belavkin--Kolokoltsov formulas with $\lambda=1$ and with the initial
data $\widetilde\Phi_0$ being equal to $\Phi_0$ scaled in $\lambda$ times (the
least equality follows from the scaling homogenity of the
Euler--Belavkin--Kolkol'tsov formulas). As a sequence of results of
(Kolokoltsov 1993) (the
dependence of the $q_R$--affine current $J$ on $u$ is not essential in view of
the translation invariance) the expression
$\limsup_{t\to\infty}E_{\widetilde\Phi_0}D_J(\widetilde\Phi(t,[\omega]))$,
being the Kolokoltsov coefficient $\kappa(A_t,J)$ for the pair $(A_t,J)$,
does not depend on $\widetilde\Phi_0$, and hence, it is certainly independent
on $\lambda$.
\enddemo

\remark{Remark 6} The sketch of the proof is rather instructive itself. Instead
of difficult calculations of the stationary probability measure (cf.
Kolokoltsov 1993, see also Huang {\it et al\/} 1983) and a complicated
estimation of its $\lambda$--behaviour (that is non--trivial to perform rather
in the simplest 2--dimensional case considered in (Kolokoltsov 1993)) we use
general group--theoretical properties (the translation invariance and the
scaling homogenity) of the Euler--Belavkin--Kolokoltsov formulas, combining
them with the strong results of (Kolokoltsov 1993) on an existence of the
Kolokoltsov coefficient $K=\kappa(A_t,J)$ and its independence on the initial
data.
\endremark

\remark{Comments on the proof} Concerning the sketch of the proof two remarks
on some details should be made. First, in view of the dependence of the
angular field $A_t(u,\dot u)$ on the controlling parameters the unique
stationary probability measure does not exist; however, we consider all
controlling parameters as slow ones so one may {\sl assume} that there exists
the slowly evoluting stationary probability measure, which form depends only
on the current values of controlling parameters (of course, it is clear that
such assumption is natural from mathematical physics point of view, however,
it means a certain {\it "gap"\/} in the rigorous proof from pure mathematics
one; but here any "purification" will be out of place). Such parameters varies
through a compact set (in the continuous version, or may have only finite
number of values in the lattice version), so one can define the Kolokoltsov
coefficient as the supremum of such coefficients calculated for the measures
from the compact (or finite) set (just this circumstance causes the appearing
of "$\limsup$" in Lemma 1). However, second, now one may use the scaling
rigorously only for infinite regions, whereas we have to deal with finite ones
(the screen of a display or clusters); however, the transition to the compact
regions may only cause that the Kolokoltsov coefficient $K$ being a function
of $\lambda$ decreases if $\lambda$ tends to infinity.
\endremark

Let's $Q$ be the coordinate operator $Qf(x)=xf(x)$; $J^\circ$ be a singular
part of the current $J$ (Juriev 1992, 1994a), i.e. $J^\circ(u)=(Q-u)^{-1}$.

\proclaim{Lemma 2}
$$E_{\Phi_0}
\left(D_J(\Phi(t,[\omega]))-D_{J^\circ}(\Phi(t,[\omega]))\right)
\rightrightarrows 0\quad\text{ if }\quad q_R\to 0. $$
\endproclaim

It should be marked that the statement of the lemma na\"\i vely holds only in
the continuous version; after a finite--dimensional approximation the
expression "$\rightrightarrows 0$" should be understand as the l.h.s. becomes
uniformely less than a sufficiently small constant $\epsilon$ (which depends
on the chosen approximation), when $q_R$ tends to zero.

\demo{Hint to the proof} The lemma follows from the explicit computations of
eigenfunctions of a $q_R$--conformal current $J(u)$.
\enddemo

\proclaim{Main Theorem}
$$(\forall\Phi_0)\quad\lim_{\lambda\to\infty, q_R\to 0}\limsup_{t\to\infty}
E_{\Phi_0} D_Q(\Phi(t,[\omega]))=0.$$
\endproclaim

The statement of the theorem is a natural sequence of two lemmas above; it
remains true in the multi--user mode (Juriev 1994d) also. Certainly, the
statement of the
theorem na\"\i vely holds only in the continuous version (cf. Lemma 2); after
a finite--dimensional approximation the equality of the limit to 0 should mean
that this limit is less than a sufficiently small constant $\epsilon$, which
depends on the chosen approximation.

\remark{Comment} Thus, we received that the Belavkin--Kolokoltsov watch--dog
effects in stochastic mobilevision appear for all values of the accuracy of
measurement $\lambda$ for sufficiently small values of parameter $q_R$.
Moreover, if $\lambda$ increases and $q_R$ tends to $0$ the stochastic cores
may be diminished to several pixels.
\endremark

\remark{Remark 7} Note that the Belavkin--Kolokoltsov watch--dog effects
appear only in the models of SMV with sufficiently small values of the
internal parameter $q_R$, which measures the localization of interaction
($q_R=0$ mens the local case). However, $q_R$, being an internal parameter,
may be chosen in arbitrary way, so the condition $q_R\ll 1$ may be always
provided.
\endremark

\head IV. CONCLUSIONS\linebreak (SUMMARY OF RESULTS)
\endhead

Thus, the results may be briefly summarized.

First, let's emphasize once more that the main difference of SMV from the
ordinary MV is that the stochastization of eye tremor in the first is
considered as an internal process, so its amplitude characteristics may be
{\it reinforced}. Second, for {\it all values\/} of $\lambda$ (a free parameter
of such stochastization, which measures the reinforcing of the amplitude of
tremor --- the so--called accuracy of measurement) the Belavkin--Kolkoltsov
watch--dog effects for stochastic dynamics of image in SMV are observed (it
means that stochastic cores of image have finite sizes for all times) for
sufficiently small values of an additional internal parameter $q_R$; it
confirms the presence of watch--dog effects also in the models of ordinary MV
with the same $q_R$. Moreover,
third, if the value of $\lambda$ is great enough, whereas $q_R\ll 1$ than the
stochastic scores of
SMV image may be diminish to several pixels. Such effect, which is produced by
the reinforcing of $\lambda$, may be effectively used in practical
computer-graphics for various purposes as it was marked in the introduction.
Some further discussions of significance of the obtained results for other
applications may be found in the next paragraph.

\head V. REMARKS ON APPLICATIONS, THEIR RELATIONS TO OBTAINED RESULTS AND
GENERALIZATIONS
\endhead

\remark{Remarks on applications} Besides theoretical importance for the
interactive visualization of 2D quantum field theory the results of the paper
seems to be useful for applications to ({\bf 1}) the elaboration of
computer-graphic interactive systems for psychophysiological
self--regulation and cognitive stimulation (Juriev 1994b,c), ({\bf 2}) the
interactive computer-graphic modelling of a "quantum computer" (Juriev
1994c) (see (Deutsch 1985, Josza 1991, Deutsch \&\ Josza 1992) for a general
discussion on "quantum computers" and their use for rapid computations as well
as (Unruh 1994) on fundamental difficulties to construct the "physical"
non-interactive "quantum computer"), which may be used for an actual problem
of the accelerated processing of the complex sensorial data in the "virtual
reality" (visual--sensorial) networks, ({\bf 3}) the creation of
computer graphic networks of tele\ae sthetic communication (Juriev 1994c).
\endremark

Let's discuss a significance of obtained results for these applications.

\remark{Comment: Obtained results and applications}

({\bf 2}) is directly related to our results because the maintaining of the
coherence is the main problem for "quantum computers". As it was mentioned
earlier (Juriev 1994c) MV may be regarded as an interactive computer-graphic
simulation of a "quantum computer" behavior. The presence of free parameters
(such as $\lambda$) in SMV allows to maintain the coherence for long times with
an arbitrary precision in the interactive mode.

Moreover, such interactive computer-graphic simulations may be more useful
than the original "quantum computers" for the "virtual reality" problems in
view of the implicit presence of graphical datain the interactive mode. A
reorganization of these data by the secondary image synthesis (Juriev 1994e)
and their representation via MV or SMV may allow an accelerated parallel
processing of the complex sensorial data in such systems.

({\bf 1}) and ({\bf 3}) are indirectly related to our results because they
depend on a solution of the main problem of dynamics in interactive
psychoinformation computer-graphic systems (a problem of the nondemolition of
images). For ({\bf 3}) its solution allows to transmit the graphically
organized information without a dissipation and additional errors. For ({\bf
1}) its solution allows to consider a long--time self--organizing interactive
processes, which play a crucial role in systems for psychophysiological
self--regulation and cognitive stimulation.
\endremark

So it should be stressed that the obtained results are essential for the
prescribed applications.

Now let's discuss the possible generalizations.

\remark{Remarks on generalizations and perspectives} Really one consider a
random (discrete) simulation of the continuous Brownian motion and stochastic
differentials.  It may be rather interesting to replace it by any their
perturbation (f.e. by some version of the weakly self--avoiding or
self--attracting walks, especially by their finite memory approximations).
\endremark

First, these generalizations are motivated by the fact that Brownian motion
may be not the best stochastization of the eye tremor. Really, it may be
considered only as a first approximation for tremor, whereas the more
complicated models will be preferable. However, it seems that the watch--dog
effects are conserved by any form of the weakly self--attracting
perturbations, which are the most realistic candidates for tremor.

Second, it seems to be rather interesting to use the decoupling of
high--frequency tremor from slow eye movements (including saccads) and an
internal character of its stochastic simulations for the organization of
various "intelligent" forms of man--machine interaction (the so--called {\it
"semi--artificial intelligence"\/}). In such approach the stochastized tremor
plays a role of an internal observer (cf. R\"ossler 1987), which presence is
crucial for a self--organization of graphical data in systems of the
semi--artificial intelligence (Kaneko \&\ Tsuda 1994). But this topic (though
being related to ({\bf 1}) above) seems to be too manysided and too intriguing
that this paper is not a suitable place to discuss it further.

\head VI. ACKNOWLEDGEMENTS
\endhead

The author is undebtful to Prof.Dr.V.N.Kolokoltsov (Applied Mathematics
Department, Moscow Institute for Electronics and Mathematics (MIEM), Moscow,
Russia) for an attention and numerous discussions as well as to Laboratoire de
Physique Th\'eorique de l'\'Ecole Normale Sup\'erieure for a kind hospitality
as well as for a soft and delicate intellectual atmosphere during a work on
the paper.

The author also thanks his referees for very valuable comments, constructive
remarks and useful suggestions.

\head References \endhead

Alekseev, A., Faddeev, L., Semenov--Tian--Shansky, M. \&\ Volkov, A. 1991
The unravelling of the quantum group structure in the WZNW theory.
{\it Preprint\/} CERN 5981/91.

Alekseev, A., Faddeev, L., \&\ Semenov--Tian--Shansky, M. 1992 Hidden quantum
groups
inside Kac--Moody algebra. {\it Commun.~Math.~Phys.\/} {\bf 149}, 335-345.

Belavkin, V.P. 1988 Nondemolition measurements, nonlinear filtering and
dynamic programming of quantum stochastic processes. {\it Lect.~Notes
Contr.~Inform.~Sci.\/} {\bf 121}

Belavkin, V.P. \&\ Kolokoltsov, V.N. 1991 Quasiclassical asymptotics of
quantum stochastic equations. {\it Theor.~Math.~Phys.\/} {\bf 89}, 1127-1138.

Chiu, C.B., Sudarshan E.C.G. \&\ Misra, B. 1977 Time evolution of unstable
quantum states and a resolution of Zeno's paradox. {\it Phys.~Rev.~D\/} {\bf
16}, 520-529.

Deutsch, D. 1985 Quantum theory, the Church--Turing principle and the
universal quantum computer, {\it Proc.~Roy.~Soc.~A\/} {\bf 400}, 97-117.

Deutsch, D. \&\ Jozsa, R. 1992 Rapid solution of problems by quantum
computation. {\it Proc.~Roy.~Soc.~A\/} {\bf 439}, 553-558.

Dynkin, E.B. 1965 {\it Markov processes}. Springer--Verlag.

Gihman, I.I. \&\ Skorohod, A.V. 1979 {\it The theory of stochastic
processes. III}. Springer--Verlag.

Home, D. \&\ Whitaker, M.A.B. 1992 A critical re--examination of the quantum
Zeno paradox. {\it J.~Phys.~A:~Math.~Gen.\/} {\bf 25} 657-664.

Huang, G.M., Tarn, T.J. \&\ Clark, J.W. 1983 On the controllability of
quantum--mechanical systems. {\it J.~Math.~Phys.\/}
{\bf 24}, 2608-2618.

Jozsa, R. 1991 Characterizing classes of functions computable by quantum
parallelism. {\it Proc.~Roy.~Soc.~A}. {\bf 435}, 563-574.

Juriev, D. 1992 Quantum projective field theory: quantum--field analogs of
Euler formulas. {\it Theor.~Math.~Phys.\/} {\bf 92}, 814-816.

Juriev, D. 1994a Quantum projective field theory: quantum--field analogs of
Euler--Arnold equations in projective $G$--hypermultiplets. {\it
Theor.~Math.~Phys.\/} {\bf 98}, 147-161.

Juriev, D. 1994b Octonions and binocular mobilevision, {\it E-print\/} (LANL
archive on Theor.~High Energy Phys.): {\it hep-th/9401047}.

Juriev, D. 1994c Visualizing 2D quantum field theory: geometry and
informatics of mobilevision, {\it E-print\/} (LANL archive on Theor.~High
Energy Phys.): {\it hep-th/9401067}.

Juriev, D. 1994d The advantage of a multi--user mode, {\it E-print\/} (LANL
archive on Theor.~High Energy Phys.): {\it hep-th/9404137}.

Juriev, D. 1994e Secondary image synthesis in electronic computer photography,
{\it E-print\/} (LANL archive on Adap.~Self-Org.~Stoch.): {\it
adap-org/9409002\/} [in Russian].

Kaneko, K. \&\ Tsuda, I. 1994 Constructive complexity and artificial reality:
an
introduction, {\it E-print\/} (LANL archive on Adap.~Self-Org.~Stoch.): {\it
adap-org/9407001}.

Kolokoltsov, V.N. 1991 Application of the quasiclassical methods to the
investigation of the Belavkin quantum filtering equation. {\it Math.~Notes\/}
{\bf 50}, 1204-1206.

Kolokoltsov, V.N. 1993 Long time behavior of continuously
observed and controlled quantum systems (a study of the Belavkin quantum
filtering equation). {\it Preprint\/} Institut f\"ur Mathematik,
Ruhr-Universit\"at-Bochum, No.204.

Misra, B. \&\ Sudarshan, E.C.G. 1977 The Zeno's paradox in quantum theory.
{\it J.~Math.~Phys.\/} {\bf 18}, 756-763.

Peres, A. 1993 {\it Quantum theory: concepts and methods}. Dordrecht, Boston:
Kluwer Acad. Publ.

R\"ossler, O.E. 1987 Endophysics. In {\it "Real brains, artificial minds"\/}
Eds. J.L.~Car\-ti and A.~Karlqvist, North Holland.

Skorohod, A.V. 1982 Operator stochastic differential equations and stochastic
semigroups. {\it Russian Math.~Surveys\/} {\bf 37:6}, 177-204.

Unruh, W.G. 1994 Maintaining coherence in quantum computers, {\it E-print\/}
(LANL archive on Theor.~High Energy Phys.): {\it hep-th/9406058}.
\enddocument